\documentclass[aps,apl,reprint]{revtex4-1}
\usepackage{amsfonts}
\usepackage{amsmath}
\usepackage{amssymb}
\usepackage{bm}
\usepackage{graphicx}
\usepackage[T1]{fontenc}
\begin{document}

\title{Third Harmonic THz Generation from Graphene in a Parallel-Plate Waveguide} 

\author{Parvin Navaeipour}
\affiliation{Department of Physics, Engineering Physics and Astronomy,Queen's University, Kingston, Ontario K7L 3N6, Canada}

\author{Ibraheem Al-Naib}

\affiliation{ Biomedical Engineering Department, College of Engineering, University of Dammam, Dammam 31441, Saudi Arabia}

\author{Marc M. Dignam}
\affiliation{Department of Physics, Engineering Physics and Astronomy,Queen's University, Kingston, Ontario K7L 3N6, Canada}

\email{p.navaeipour@queensu.ca}

\begin{abstract}
Graphene as a zero-bandgap two-dimensional semiconductor with a linear electron band dispersion near the Dirac
points has the potential to exhibit very interesting nonlinear optical properties. In particular, third harmonic
generation of terahertz radiation should occur due to the nonlinear relationship between the crystal
momentum and the current density. In this work, we investigate the terahertz nonlinear response of graphene inside a parallel-plate waveguide. We optimize the plate separation and Fermi energy of the graphene to maximize third harmonic generation, by maximizing the nonlinear interaction while minimizing the loss and phase mismatch. The results obtained show an increase by more than a factor of $100$ in the power efficiency relative to a normal-incidence configuration for a $2$ terahertz incident field.
\end{abstract}

\maketitle

Graphene as a 2D allotrope of carbon has the potential to exhibit very interesting nonlinear optical properties. The linear dispersion relation of the electrons near the Dirac points leads to a constant electron speed \cite{wallace,sarma,Mak}. Thus, the intraband current induced in the graphene by terahertz (THz) fields displays clipping as we increase the amplitude of the incident field, which gives rise to third and higher harmonics in the current and transmitted electric field \cite{mikailov,Mikhailov2010,Bowlan2014,Alnaib2015,Ibra2015}. THz radiation has applications in areas including ultrahigh speed wireless communications, medical imaging and sensing, high speed computing, and optics \cite{thz,tonouchi,tuniz}. However, current technologies for generating THz radiation are limited. Exploiting the nonlinear response of graphene enables one to produce higher-frequency THz radiation through the generation of harmonics. There have been several experimental efforts examining harmonic generation in graphene with the radiation \emph{normally} incident upon the graphene sheet \cite{Bowlan2014,Hafez,Paul,PBowlan}. However, in this work, we consider a configuration in which the radiation is incident \emph{parallel} to the plane of the graphene, which is located inside a waveguide. This increases the interaction time between the radiation and graphene and thereby potentially increases the generated third harmonic field.\par
 We employ a Parallel-Plate Waveguide(PPW) rather than a dielectric waveguide because it allows stronger confinement of the THz field and thereby yields a stronger interaction with the graphene. There are two difficulties that need to be overcome in employing this waveguide geometry. First, the linear conductivity of the graphene can result in high losses at the fundamental and third harmonic as they propagate down the waveguide. Second, for propagation over distances of more than a few hundred microns, phase mismatch between the fundamental and the third harmonic can become a significant problem. As we shall show in this work, through optimization of the graphene Fermi energy, the plate separation, and the structure length, it is possible to overcome both of these difficulties and to increase the power conversion efficiency by more than a factor of 100 over that attainable in the conventional normal-incident configuration. \par 
\begin{figure}
  \centering
  \includegraphics[width=0.8 \columnwidth]{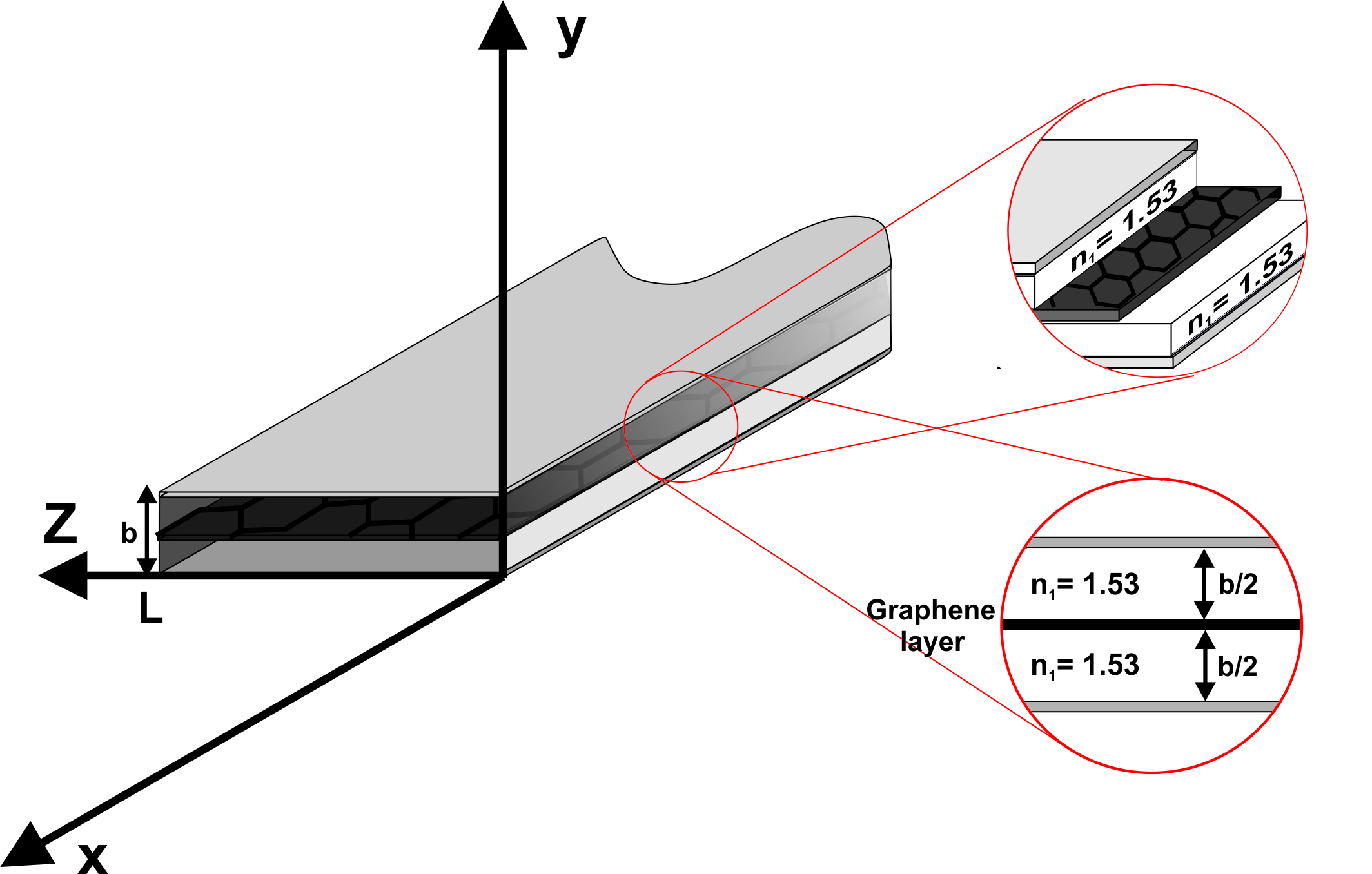} 
  \caption{Metallic parallel-plate waveguide with graphene inside. The inner material of the waveguide is Polyolefin and the graphene is placed at the center of the waveguide.}
  \label{fig:Figure1}
\end{figure}\par
Our parallel-plate waveguide consists of two metallic plates placed at $y= 0$ and $y= b,$ with the graphene midway between the plates as shown in Fig. 1. We choose the inner material of the waveguide to be cyclic Polyolefin, with a refractive index of $ n_1 = 1.53$ due to its compatibility with graphene and ease of fabrication \cite{Nestor}. The THz wave propagates in the $z-$direction and we take the plates to be perfect conductors that are infinite in the $x-$direction. Using Maxwell's equations and considering boundary conditions between the regions below and above the graphene, we solve for the linear electric field in the waveguide \cite{Pozar} for a monochromatic field of frequency, $\omega$. As the fundamental mode propagates along the waveguide, the interaction of the electric field with the graphene generates a nonlinear current in the graphene due to its third order conductivity, $\sigma^{(3)} (\omega)$. We employ a Green function approach to solve for the generated third harmonic electric field in the PPW, including self-consistently the ohmic losses in the graphene at the fundamental and third harmonic. We optimize the plate separation and Fermi energy of graphene to obtain phase matching and low loss, and thereby maximize third harmonic generation.\par
For waves travelling in the $+z$ direction, the linear electric field below and above graphene for the $n^{th}$ transverse electric $(TE)$ mode can be expressed as
\begin{align}
E_x^{(1)} (y,z;\omega) =& {E}_n e^{ i \tilde{\beta}_n(\omega) z} \sin(\tilde{k}_ny)      \quad \quad    0 <y <b/2 \\
E_x^{(1)} (y,z;\omega) =& -{E}_n e^{ i \tilde{\beta}_n(\omega) z} \sin(\tilde{k}_n(y-b)),    \quad    b/2 <y <b \nonumber
\end{align}
where $E_n$ is the amplitude of the field at $z=0$, and $\tilde{k}_n$ is the complex wavenumber for the field's $y-$ dependence, which depends on the linear conductivity of the graphene (see Sec. \textrm{I} in the supplemental material). The complex propagation constant in the $z-$direction, $\tilde{\beta}_n$, of the $TE_n$ mode is given by
\begin{equation}
\tilde{\beta}_n (\omega)= \sqrt{\bigg(\dfrac{n_1 \omega}{c_0}\bigg)^2 - \tilde{k}_n^2},
\end{equation}
where $c_0$ is the speed of light in vacuum. If there is no graphene, \textit{i.e}., for a \emph{bare waveguide}, $ \tilde{k}_n \equiv k_n^0 =n \pi/b$ and $\tilde{\beta}_n \equiv \beta_n^0$, where $n$ is an integer.\par  
In this work, we take the input field at $\omega$ to be in the $ n=1$ mode and neglect the depletion of the pump due to the nonlinear interaction, but include linear loss. Using a Green function approach, we calculate the generated third harmonic electric field (see Sec. \textrm{II} in the supplemental material for more details). For a finite length of graphene extending from $0$ to $L$, the third order electric field at $(y,z)$ (where $0\leq z\leq L$) is given by 
\begin{align}
E_x^{(3)} (y,z;3\omega) =& \int_{0}^{L} G(z,z_0) J_0^{tot} (z_0;3\omega) d z_0,
\end{align}
where $G(z,z_0)$ is the Green function, which can be expanded exactly in terms of the \emph{bare waveguide} modes as 
\begin{equation}
G(z,z_0) = \sum_{n = 1}^{\infty} \dfrac{-3 \omega \mu}{{\beta}^0_n(3\omega) b} \sin(\dfrac{n \pi }{2}) \sin(\dfrac{n \pi y}{b})e^{i \beta_n^0(3\omega) \vert z-z_0 \vert} 
,\end{equation}
where $\mu$ is the permeability of the dielectric and $ J_0^{tot} (z_0; 3\omega)$ is the total nonlinear current density at $3 \omega $ in the graphene, which given by 
\begin{equation}
J_0^{tot} (z; 3\omega) = \sigma^{(3)} (\omega) [E_x^{(1)} (\dfrac{b}{2},z;\omega)]^3 + \sigma^{(1)} (3 \omega) E_x^{(3)} (\dfrac{b}{2},z;3\omega). 
\end{equation}
There are two contributions to the nonlinear current. The first term arises from the nonlinear conductivity of graphene at $ \omega$. In this work, we use the theoretical expression for $\sigma^{(3)}(\omega)$ derived by Cheng et al. \cite{cheng}. The dependence of the conductivity on the Fermi energy for a fixed frequency of $\omega_0/(2\pi) = 2.0  \,\:THz $ is shown in Fig. S2 in the supplementary material. The second term in the nonlinear current density is the \textit{self-interaction current}, which arises from the linear interaction of the generated third harmonic electric field with graphene via the linear conductivity, $\sigma^{(1)} (3\omega) $, of the graphene at $ 3 \omega $. For the doped graphene that we consider here, at THz frequencies the linear conductivity is given simply by the Drude model for graphene \cite{Ibra2015,horng} (see Eqs. (S14) and (S15) in the supplemental material). We note that, for the range of Fermi energies considered in this work, as the Fermi energy is decreased, the nonlinear conductivity increases, while the linear conductivity decreases. Thus, one expects that a lower Fermi energy will result in a larger third harmonic field. However due to difficulties in achieving a uniform doping over graphene sheets that are millimeters in length, achieving Fermi energies below $20 \, meV$ is very challenging and so we set this as a lower limit.\par
Because the third harmonic field appears on both sides, we need to solve Eq. (3) self-consistently. In Sec. \textrm{IV} of the supplemental material, we show how this is done by converting the integral to a sum and inverting the resulting matrix. In Fig. 2, we plot the third harmonic electric field at the graphene calculated using this approach when we include only one mode ($n = 1 $) in the Green function expansion of Eq. (4). We take the waveguide plate separation to be $70 \, \mu m$, the fundamental frequency to be $2.0 \, \:THz$, and the input electric field to be $ 5 \, kV/cm$. The dashed black curve in Fig. 2 is the generated third harmonic electric field without the self-interaction term. Initially the third harmonic grows, then it oscillates until it settles down to a field of about $0.014 \, kV/cm$. The initial growth is due, of course, to the current in the graphene at the third harmonic. The oscillations arise from the poor phase matching between the fields at $\omega_o$  and $3\omega_o$, which are both in the $n=1$ mode. For this waveguide, the beat length between these two fields is $ 0.12 \, mm$, which is what gives the period of the observed oscillations. The decay of these oscillations is due to the exponential reduction in the amplitude of the fundamental field as it is absorbed linearly by the graphene. The solid blue curve is the third harmonic field calculated with the inclusion of the self-interaction term; as can be seen, the self-interaction results in a decay in the third harmonic, but at a much slower rate than for the fundamental, due to the decrease in the linear conductivity with frequency. 
\begin{figure}[h]
  \centering
  \includegraphics[width=0.8 \columnwidth]{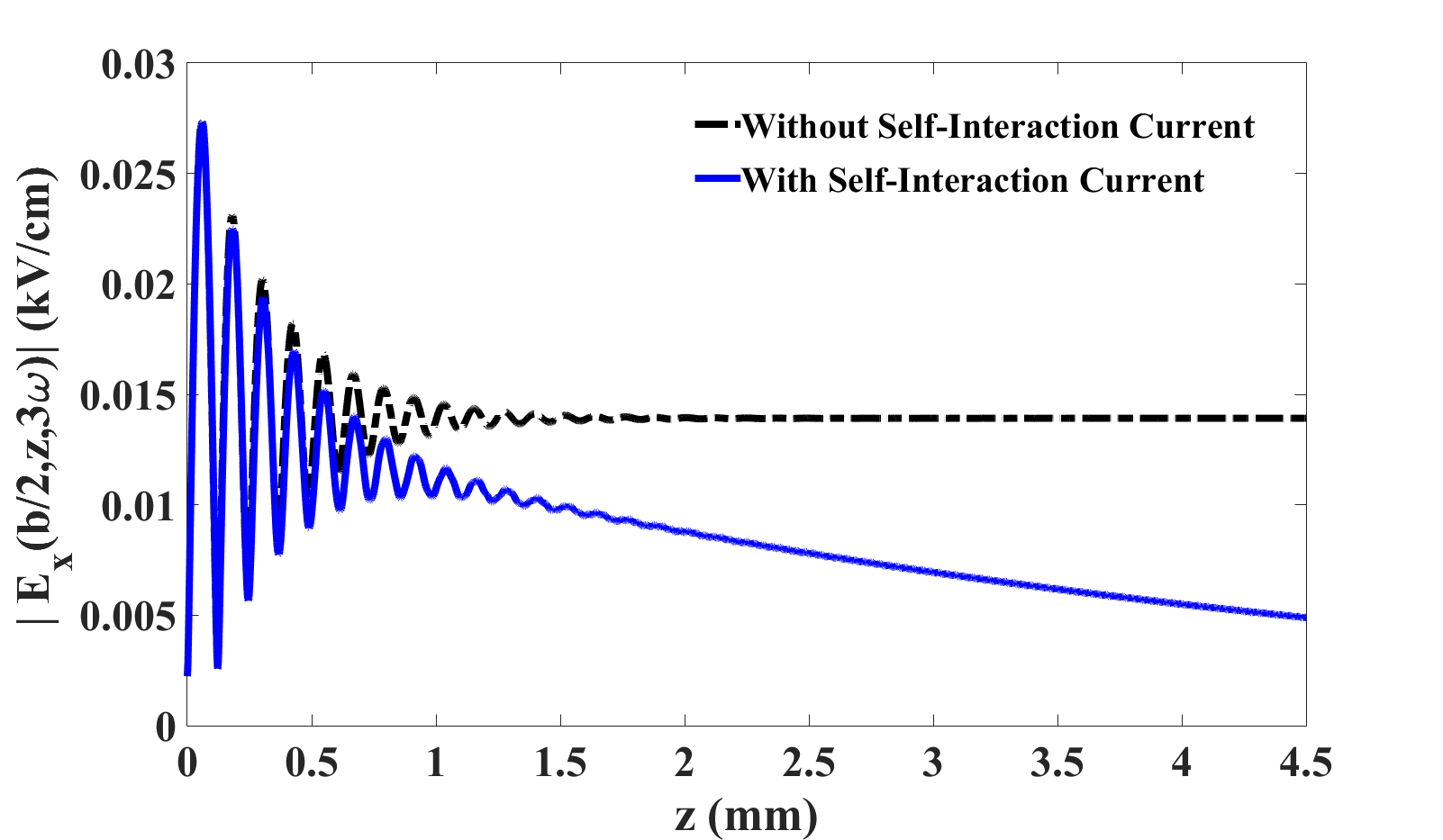} 
  \caption{The third harmonic field at the graphene for an input field of $ 5 \, kV/cm$ when only one mode $(n=1)$ is included in the Green function expansion. The dashed black curve (solid blue curve) is for the calculation when the self-interaction is (is not) included. The plate separation is $ b = 70 \, \mu m $ and the Fermi energy is $E_f = 50 \, meV$. } 
  \label{fig:Figure5}
\end{figure}\par 
If we wish to generate a strong third harmonic, we need to include more modes in our Green function expansion \emph{and} ensure that there is good phase matching between the fundamental in the $TE_1$ mode and the third harmonic in at least one of the $TE_n$ modes. This will occur if the effective refractive index difference between these modes,  
\begin{equation}
\Delta_n \equiv n_{eff}^{n}(3 \omega_0) - n_{eff}^{n=1} (\omega_0),
\end{equation}
is close to zero, where, $n_{eff}^{n}( \omega)$ is the effective refractive index for $ n^{th}$ mode at $ \omega$ defined as $n_{eff}^{n}( \omega) \equiv {\Re  \lbrace \tilde{\beta}_n(\omega) \rbrace}/k $, where $ k \equiv \omega/c_0$. For a \emph{bare waveguide}, using Eq. (2), it is easily seen that Eq. (7) is exactly satisfied for $n = 3$. We therefore choose a device that has three and only three propagating modes at $3\omega_0$. This constrains plate separation to be $49 \, \mu m \leq b \leq 70 \, \mu m$. There is no interaction between of the field in the  even modes and the graphene, so we ignore all even modes in our analysis. \par 
In Fig. 3, we plot (a) $\Delta_3$  and (b) the field loss coefficient, $\alpha \equiv \Im (\tilde{\beta}_n (\omega))$, at $\omega_o$ for three different Fermi energies as a function of plate separation. As expected, as the Fermi energy decreases, so generally does the phase-mismatch and the loss. For $ E_f = 50 \, meV $, for example, perfect phase matching occurs for $ b = 49.64 \, \mu m$. However, for this plate separation, the frequency of the propagating mode is close to the cutoff frequency, which yields a huge loss in the waveguide, as can be seen in Fig. 3(b). In addition, it is easily shown that the group velocity dispersion is very large near cutoff. Therefore, to minimize phase mismatch without suffering these two detrimental effects, one should choose the plate separation to be $b = 70 \, \mu m$.
\begin{figure}[h]
  \centering
  \includegraphics[width=0.9 \columnwidth]{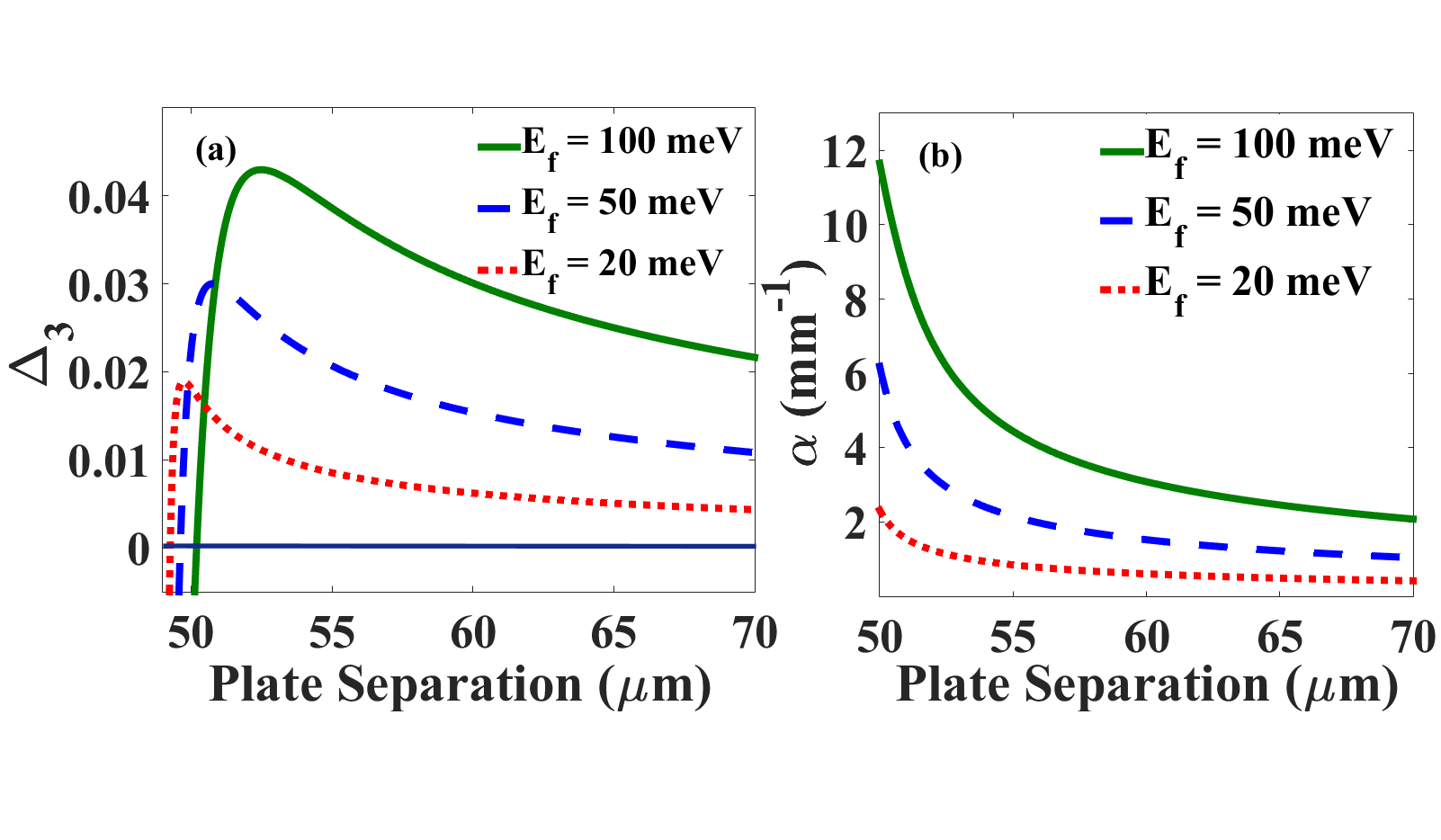} 
  \caption{(a) Effective refractive index difference between the $n=3$ mode at $ 3 \omega_0$ and $n=1$ mode at $ \omega_0$ and (b) Loss coefficient in the $n=1$ mode at $ \omega_0$ as a function of plate separation for three different Fermi energies.}
   \label{fig:Figure6}
  \end{figure} 
For a Fermi energy of $50 \, meV$ and $b = 70 \, \mu m$, the loss coefficient at $\omega_o$ is $1.04 \, mm^{-1}$, which gives the decay length of the oscillations seen in Fig. 2. We note that for all of the parameters and frequencies considered in this work, the loss coefficient due to the graphene is more than an order of magnitude larger than that which would arise due to the finite conductivity of the gold plates; this justifies our approach of taking the plates to be perfect conductors.\par
The third harmonic field at the graphene for $ b = 70 \, \mu m $ and a Fermi energy of $50 \, meV$ is plotted in Fig. 4(a), where the expression in Eq. (4) has been summed to convergence \cite{Convergenote}. The black dashed curve gives the result \emph{without} the self-interaction. As can be seen there is a rapid rise in the generated field and the final value is much higher than was found when only the $n=1$ mode was taken into account (see Fig. 2). The oscillations arise primarily due to the phase mismatch between the third harmonic field in the $n=1$ and $ n= 3$ modes, which is why they persist even when the fundamental is essentially gone ($z>1 \, mm$). When the self-interaction is included (solid blue curve), the peak field is somewhat reduced and the field decays as it propagates. For this waveguide, the maximum third harmonic is reached at a length of about $0.78 \, mm$. In Fig. 4(b) we plot the evolution of the generated third harmonic field with loss included for three different Fermi energies. As can be seen, the peak field occurs sooner and decreases dramatically as the Fermi energy is increased; this is due to the increase in the linear loss, the decrease in the nonlinear conductivity (see in Fig. 3(a)), and poorer phase matching between the $n =1 $ and $ n=3$ modes as the Fermi energy is increased.\par 
\begin{figure}[h]
  \centering
  \includegraphics[width=0.9 \columnwidth]{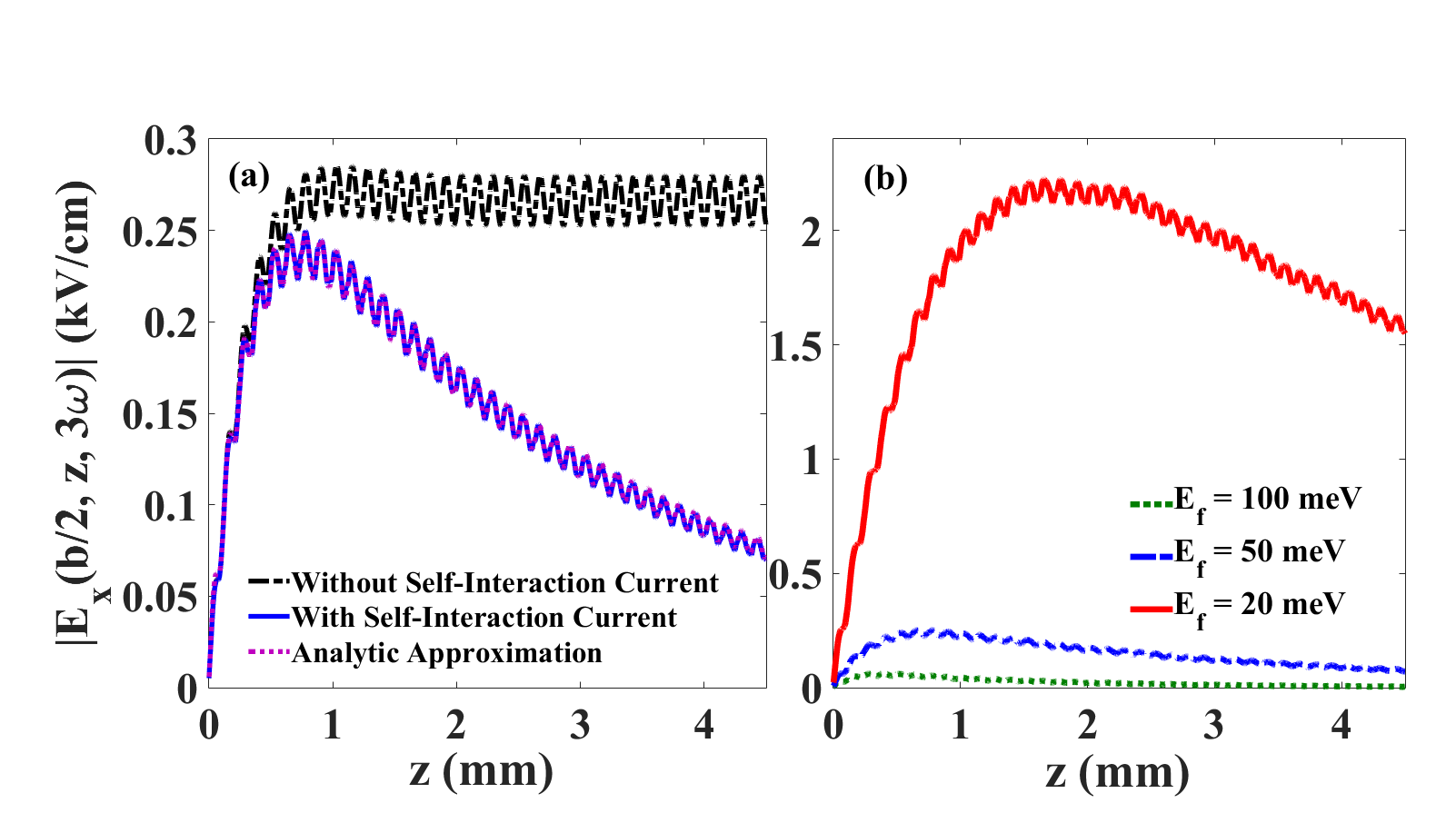} 
  \caption{ Generated third harmonic electric field at the graphene using the full Green function (a) for $E_f = 50 \, meV$ and (b) for three different Fermi energies. In both plots, $ b=70 ~\mu m$ and the amplitude of the input field is $E_1 = 5 \, kV/cm$. In (a), we plot the field without the self-interaction (dashed black), with the self-interaction (solid blue) and  using the approximate two-lossy-mode method described later (dotted red).}
  \label{fig:Figure8}
\end{figure} 
We now consider the power efficiency of the device. This is defined as the ratio of the power in the third harmonic at the end of the waveguide to the power in the fundamental at the beginning of the waveguide. In Fig. 5 we plot the maximum power efficiency as a function of the Fermi energy, for $b=70 \, \mu m$. As the efficiency simply scales as $E^{4}_1$, we only present the results for  $E_1=5 \, kV/cm$. The device length is taken to be that at which the efficiency is a maximum; the inset to Fig. 5 shows this optimum length as a function of the Fermi energy. As expected, decreasing the Fermi energy leads to higher power efficiencies. Clearly for high enough fields, our calculation will yield an unphysical power efficiency that is greater than $100\%$, which is simply the result of the undepleted pump approximation that we have employed. However, our results indicate that power efficiencies of greater than $10\%$ can be achieved for modest field amplitudes as low as $E_1 = 5 \, kV/cm$.  
\begin{figure}[h]
  \centering
  \includegraphics[width=0.9 \columnwidth]{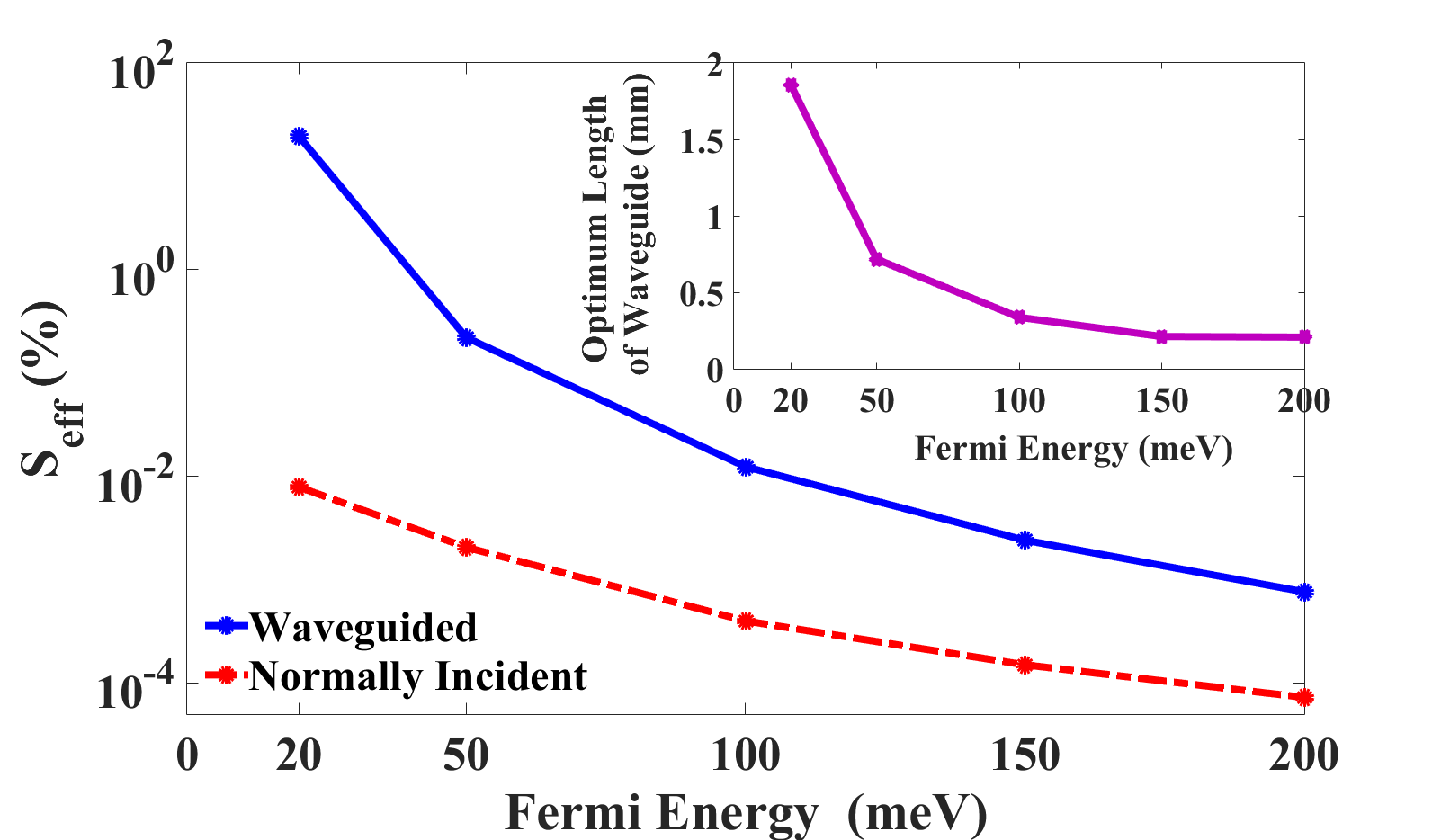} 
  \caption{Power efficiency of the waveguide system as a function of the Fermi energy for plate separation of $b = 70 \, \mu m$ and incident field amplitude, $ E_1 = 5 \, kV/cm$ is shown as solid blue curve. The dashed red curve shows the power efficiency for a normally-incident field. The inset shows the optimum length of waveguide as a function of the Fermi energy.}
  \label{fig:Figure10}
\end{figure}\par
To demonstrate the effect of the plate separation on the power efficiency, in Fig. 6, we plot dependence of the ratio of the power efficiency for plate separation, $b$, to that for $b=70 \, \mu m$ as a function of $b$. When the \emph{input electric field} is fixed at $ 5 \, kV/cm $ for all plate separations (solid blue curve), the power efficiency ratio changes relatively little with plate separation, but is a maximum for $b=70 \, \mu m$. However, if we instead take the \emph{input power} to be the same for all the plate separations (dashed red curve), then we see that the efficiency is increased significantly as the plate separation is decreased. This effect arises from the increase in the amplitude of the fundamental field at the graphene when the plate separation is decreased, as is shown in the inset to Fig. 6.  
\begin{figure}[h]
  \centering
  \includegraphics[width=0.9 \columnwidth]{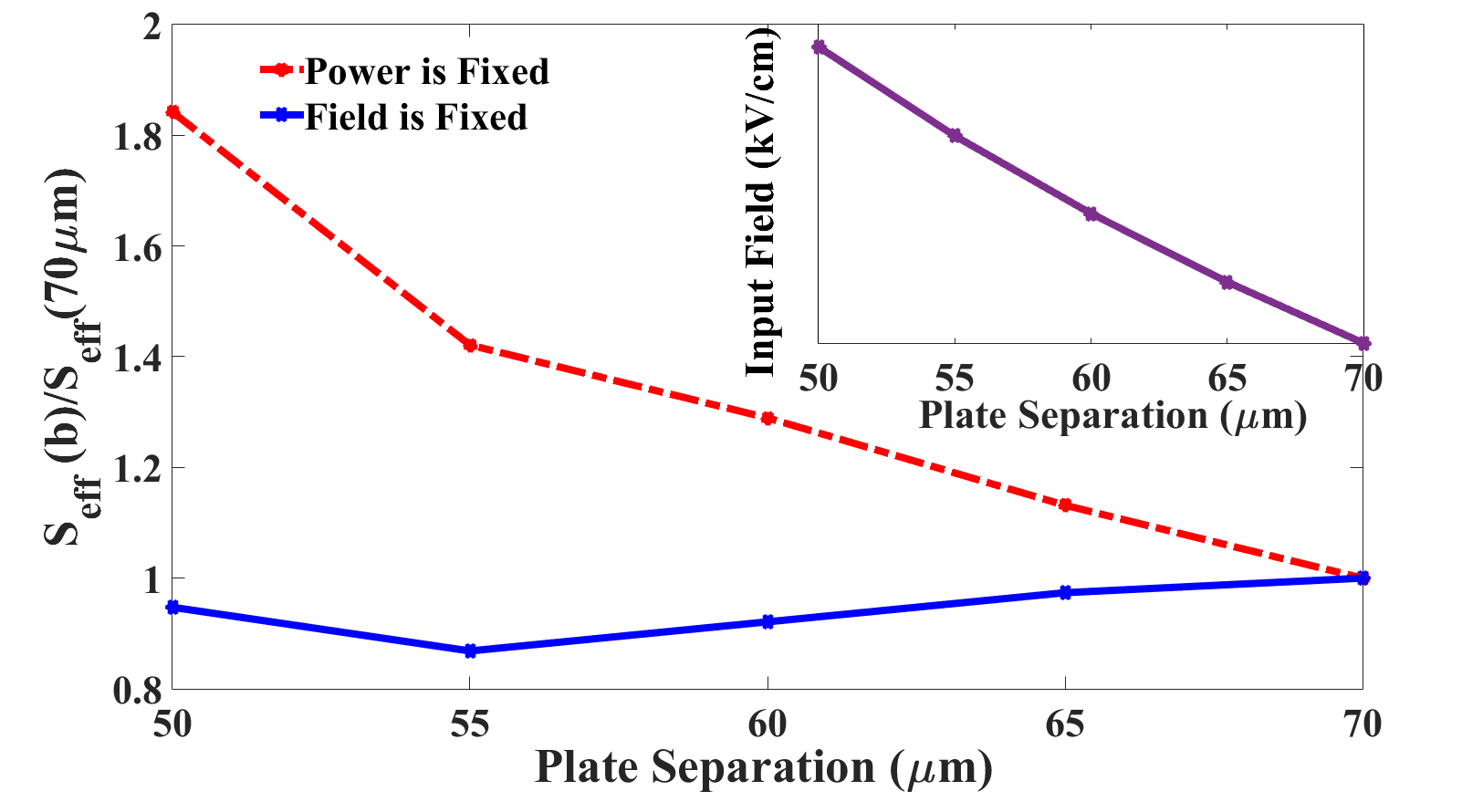} 
  \caption{Power efficiency ratio as a function of plate separation for $ E_f = 50 \, meV$. The dashed red curve shows power efficiency ratio when the input \emph{power} is fixed, while the solid blue curve shows it for a fixed input \emph{field amplitude}. The inset shows the input field amplitude at the graphene as a function of plate separation when the input power is fixed.}
  \label{fig:Figure11}
\end{figure} \par
Most previous experimental work \cite{Bowlan2014,Hafez,Paul,PBowlan} on third harmonic generation in graphene has employed a geometry in which radiation is normally incident up on a graphene sheet. We have compared the results obtained for the power efficiency for such a geometry to those obtained using our PPW geometry and we find that for $b = 70 \, \mu m$, $ E_f = 50 \, meV$, and $E_1 = 10 \, kV/cm $, the power efficiency increases from $0.033 \%$ to $ 3.5 \%$ and for $ E_f = 20 \, meV$ and $E_1 = 5 \, kV/cm $, it increases from $0.032 \%$ to $ 19.0 \%$. As can be seen in Fig. 5, for Fermi Energies below $50 \, meV$, the power efficiency in our PPW geometry is more than two orders of magnitude greater than in normally-incident geometry. \par
We finish this paper by presenting an efficient analytic approximate method for calculating the generated third harmonic field for our parallel-plate structure. In this approach, we only include the $n =1 $ and $ n=3$ modes for the third harmonic, but rather than using the \emph{bare} waveguide modes (which form a complete set), we use a subset of the \emph{lossy} waveguide modes, which are not complete but do include the self-interaction. Using Eqs. (3)-(5) but with lossy modes, without the linear term in the current and neglecting the backward-propagating fields (see supplemental material section \textrm{VI}), we obtain 
\begin{align}
E_x^{(3)} &(y,z;3\omega) =  \sum_{n=1, 3} \dfrac{3i\omega \mu}{ \tilde{\beta}_{n}(3\omega) b} \sin(\tilde{k}_{n}^{\ast}b/2) sin(\tilde{k}_{n} y) \\
& \times  \sigma^{(3)} (\omega){[E_1 \sin(\tilde{k}_{1}b/2)]}^3 
  \dfrac{e^{i 3\tilde{\beta}_{1}(\omega)z} - e^{i \tilde{\beta}_{n}(3\omega)z}}{3\tilde{\beta}_{1} (\omega) - \tilde{\beta}_{n}(3\omega)}. \nonumber
\end{align}
The dashed red curve in Fig. 4 shows the generated third harmonic electric field for $ b = 70\, \mu m$ and $E_1 = 5 \, kV/cm$ using this method. As can be seen, the agreement with the exact numerical method is excellent. We have found that this faster and simpler approach is accurate except when the frequency is very close to the cutoff frequency; in such cases, because the lossy and lossless modes become very different, multiple modes are required for convergence. For this system, this means that our approximate approach is accurate as long as the plate separation is greater than about $55 \, \mu m$.\par
In conclusion, we have calculated the generation of the third harmonic of a $2 \,\:THz$ field due to graphene inside a parallel-plate waveguide and found that power efficiencies can be increased by more than a factor of 100 relative to the results for the normal-incidence configuration. The optimum length of these structures ranges from about 200 to 2000 microns. We found that the highest efficiency occurs for low Fermi energies and that the dependence on the plate separation is relatively modest within the range where three and only three modes are above cutoff. We believe that this structure would be an excellent platform to generate third harmonic fields and to investigate the nonlinear properties of graphene at THz frequenies.\par 
\textbf{Acknowledgements} We thank the Natural Sciences
and Engineering Research Council of Canada and Queen's University for financial
support.   


\begin{thebibliography}{9}
\bibitem{wallace} P. R. Wallace, Phys Rev \textbf{71} 622 (1946).
\bibitem{sarma} S.Das Sarma, Sh, Adam, E.H. Hwang, E. Rossi, Rev Mod Phys \textbf{83}, 407 (2011).
\bibitem{Mak} K. F. Mak, L. Ju, F. Wang, T. F. Heinz, Solid State Communications \textbf{152} 1341 (2012).
\bibitem{mikailov} S. A. Mikhailov, Euro Phys Lett \textbf{79}, 27002 (2007).
\bibitem{Mikhailov2010} S. A. Mikhailov, Phys. Rev. Lett \textbf{105}, 097401 (2010).
\bibitem{Bowlan2014} P. Bowlan, E. Martinez-Moreno, K. Reimann, T. Elsaesser, and M. Woerner, Phys. Rev. B \textbf{89}, 041408 (2014).
\bibitem{Alnaib2015} I. Al-Naib,  J. E. Sipe, and M. M. Dignam, New J. Phys \textbf{17}, 113018 (2015). 
\bibitem{Ibra2015} I. Al-Naib, M. Poschmann, M. M. Dignam, Phys. Rev. B \textbf{91}, 205407 (2015).
\bibitem{ibraheem2014} I. Al-Naib, J. E. Sipe, and M. M. Dignam, Phys. Rev. B \textbf{90}, 245423 (2014).
\bibitem{Hafez} H. A. Hafez, I. Al-Naib, K. Oguri, Y. Sekine, M. M. Dignam, A. Ibrahim, D. G. Cooke, S. Tanaka, F. Komori, H. Hibino, T. Ozaki, AIP Advances \textbf{4}, 117118 (2014). 
\bibitem{Paul} M. J. Paul, B. Lee, J. L. Wardini, Z. J. Thompson, A. D. Stickel, A. Mousavian, H. Chai, E. D. Minot, Y. S. Lee, Appl. Phys. Lett \textbf{105}, 221107 (2014). 
\bibitem{PBowlan} P. Bowlan, E. Martinez-Moreno, K. Reimann, M. Woerner, T. Elsaesser, New J. Phys \textbf{16}, 013027 (2014).  
\bibitem{Riely} R. McGouran, I. Al-Naib, and M. M. Dignam, Phys. Rev. B \textbf{94}, 235402 (2016).
\bibitem{thz} P.U. Jepsen, D.G. Cooke, M. Koch, Laser Photonics Rev \textbf{5}, 124 (2011).
\bibitem{tonouchi} M. Tonouchi, Nat. Photonics \textbf{1}, 97 (2007).
\bibitem{tuniz} A. Tuniz, K. J. Kaltenecker, B. M. Fischer, M. Walther, S. C. Fleming, A. Argyros, and B. T. Kuhlmey, Nat. Commun \textbf{4}, 2706 (2013).
\bibitem{Nestor} P. D. Cunningham, N.N. Valdes, F. A. Vallejo, L. M.  Hayden,
B. Polishak, X. Zhou,J. Luo, A. K.-Y. Jen, J. C. Williams, and R. J. Twieg, J. App. Phys \textbf{109}, 043505 (2011).
\bibitem{Pozar} D. M. Pozar, Microwave Engineering, John Wiley \& Sons, Inc, 4th ed (2011).
\bibitem{cheng} J. L. Cheng,  N. Vermeulen, J. E. Sipe, Phys. Rev. B \textbf{91}, 235 (2015).
\bibitem{horng} J. Horng, C.F. Chen, B. Geng, C. Girit, Y. Zhang, Z. Hao and H. A. B. e. al., Phys. Rev. B \textbf{83}, 165113 (2011).

\bibitem{Convergenote} We find that we need only include the $n=1,3,5$ modes to obtain convergence to within 1\% of the exact result in all cases.
\end{thebibliography}
\end{document}


\title{Third Harmonic THz Generation from Graphene in a Parallel-Plate Waveguide} 

\author{Parvin Navaeipour}
\affiliation{Department of Physics, Engineering Physics and Astronomy,Queen's University, Kingston, Ontario K7L 3N6, Canada}

\author{Ibraheem Al-Naib}
\affiliation{ Biomedical Engineering Department, College of Engineering, University of Dammam, Dammam 31441, Saudi Arabia}

\author{Marc M. Dignam}
\affiliation{Department of Physics, Engineering Physics and Astronomy,Queen's University, Kingston, Ontario K7L 3N6, Canada}

\email{p.navaeipour@queensu.ca} 

\maketitle
\section{Transverse Electric modes in the Waveguide in presence of Graphene}
In this section, we find the bound modes in the waveguide in the presence of graphene. The electric and magnetic fields of waves travelling in the $ +z $ direction are given by 
\begin{align}
\mathbf{\mathcal{E}}(y,z,t) =& \mathbf{E}(y) e^{i \tilde{\beta}_n(\omega) z -i\omega t}\\
\mathbf{\mathcal{H}}(y,z,t) =& \mathbf{H}(y) e^{i \tilde{\beta}_n(\omega) z -i\omega t}\nonumber
\end{align}
Transverse electric (TE) modes must satisfy the Helmholtz wave equation. 
\begin{align}
( \dfrac{\partial^2}{\partial y^2} + \dfrac{\partial^2}{\partial z^2} + k^2 ) \mathbf{\mathcal{E}}(y,z,t) = 0&, \\
\dfrac{\partial^2 \mathbf{E}(y)}{\partial y^2} + (k^2 - \tilde{\beta}_n^2) \mathbf{E}(y) = 0. \nonumber
\end{align}
where $k = n_1 \omega \sqrt{\mu_0 \epsilon_0} = \dfrac{n_1 \omega}{c_0}$. The solution for Eq. (2) of a waveguide with two regions, below and above the graphene, is assumed to be 
\begin{align}
\mathbf{E}_{below} (y)=&  [A \cos (\tilde{k}_n y) + B \sin(\tilde{k}_n y)] \widehat{x}, \\
\mathbf{E}_{above} (y)=&  [A^{\prime} \cos (\tilde{k}_n (y-b)) + B^{\prime} \sin(\tilde{k}_n (y-b))] \widehat{x}, \nonumber
\end{align}
where $\tilde{k}_n $ is the wavenumber, defined as $\tilde{k}_n = \sqrt{k^2 - \tilde{\beta}_n^2}$, where $\tilde{\beta}_n$ is the propagation constant which is complex in the presence of the graphene determined from $\tilde{\beta}_n = \sqrt{\bigg(\dfrac{n_1 \omega}{c_0}\bigg)^2 - \tilde{k}_n^2}$ for the transverse electrical modes. 
Considering boundary conditions of the waveguide at $ y = 0$ and $ y = b $, we obtain $A = A^{\prime} = 0$. Continuity of the electric field at the graphene satisfies $ B = - B^{\prime}$. Therefore, the linear electric field equations for waves travelling in the waveguide are given by 
\begin{align}
\mathbf{E}_{below}(y) =& \mathbf{E}_n  \sin(\tilde{k}_n y) \widehat{x} \\
\mathbf{E}_{above}(y) =&  - \mathbf{E}_n  \sin(\tilde{k}_n (y-b)) \widehat{x} \nonumber
\end{align}
The surface current density, $\mathbf{\mathcal{J}}_s (y,z,t) = \mathbf{J}_s (y) e^{i \tilde{\beta}_n(\omega)z} e^{-i \omega t} $ is related to the field at the graphene by,
\begin{equation}
\mathbf{J}_s (y=b/2) = \sigma^{(1)} (\omega) \mathbf{E}_n  \sin(\tilde{k}_n b/2), 
\end{equation} 
where $\sigma^{(1)} (\omega)$ is the linear conductivity of the graphene. We now use the boundary condition on the magnetic field at the graphene to obtain a second relationship between the current and the electric field. We employ Faraday's Law to obtain the magnetic field:
\begin{align}
\mathbf{H}_{below} (y) =& \dfrac{\tilde{\beta}_n(\omega)}{\omega \mu} E_n \sin(\tilde{k}_n y) \widehat{y} - \dfrac{i \tilde{k}_n}{\omega \mu} E_n \cos (\tilde{k}_n y) \widehat{z}, \\
\mathbf{H}_{above} (y) =& -\dfrac{\tilde{\beta}_n(\omega)}{\omega \mu} E_n \sin(\tilde{k}_n (y-b)) \widehat{y} + \dfrac{i \tilde{k}_n}{\omega \mu} E_n  \cos (\tilde{k}_n (y-b)) \widehat{z}, \nonumber
\end{align} 
where $\mu$ is the permeability of the dielectric. The surface current density at the graphene can be determined by 
\begin{align}
\mathbf{J}_s (y=b/2) =& \widehat{y} \times (\mathbf{H}_{above} - \mathbf{H}_{below}) \vert_{y=b/2}\\
=& \dfrac{2 i \tilde{k}_n}{\omega \mu} E_n \cos(\tilde{k}_n \dfrac{b}{2}) \widehat{x}. \nonumber
\end{align}
Then, using Eq. (5), we obtain 
\begin{equation}
\sigma^{(1)} (\omega) = \dfrac{4i\tilde{\phi}_n}{\omega \mu b} \cot (\tilde{\phi}_n),
\end{equation}
where $ \tilde{\phi}_n \equiv \dfrac{\tilde{k}_n b}{2} $. For the linear conductivity of the graphene, we use the Drude-like model of Sec. (\textrm{III}). Then, we can solve for the complex propagation constant for any mode at a specific frequency using Eq. (8). In Fig. S1 we plot the $TE_1$ and $TE_3$ modes for a $1 \:THz$ field in  a $70 \, \mu m$ waveguide in the presence of the graphene for different Fermi energies. The electric field for both modes is largest close to the centre of the waveguide, as expected, but due to the surface current at the graphene there is a discontinuity in the derivative of the electric field at the graphene. It can be seen in the inset that the discontinuity in the mode profile increases as Fermi energy is increased, which is because of the interaction between the electric field and the graphene.  
\begin{figure}[h]
  \centering
  \includegraphics[width=0.9 \columnwidth]{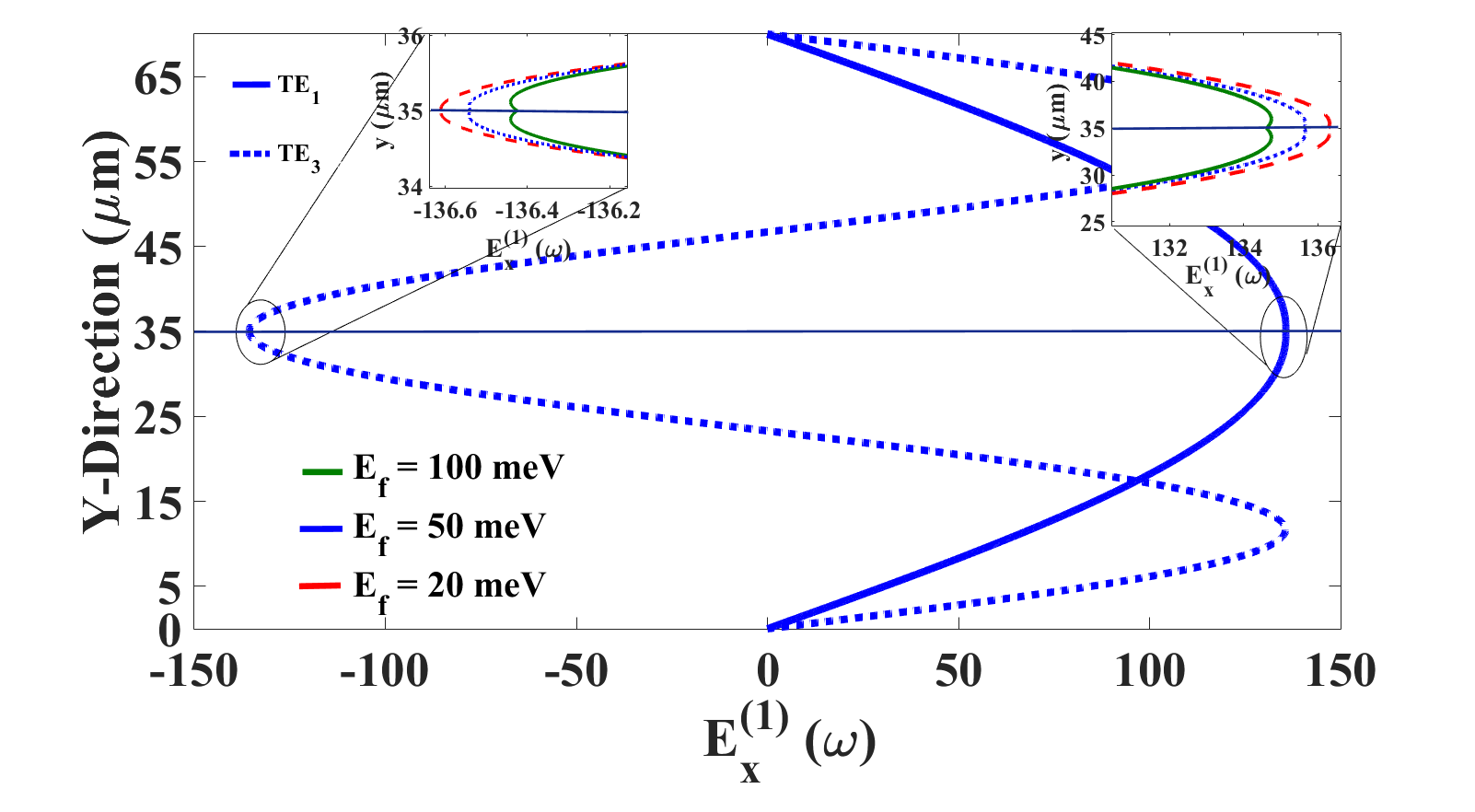}
  \renewcommand{\thefigure}{S\arabic{figure}} 
  \caption{The electric field for $ TE_1 $ and $ TE_3$ modes at $ z = 0$ travelling in a waveguide in the presence of the graphene for plate separation of $ b = 70 \,\mu m $. Dashed curve (solid curve) shows the electric field for $TE_1$ ($TE_3$). The discontinuity in the mode profile is highlighted in the insets for the $TE_1$ and $TE_3$ modes.}
  \label{fig:Figure2}
\end{figure}
\section{Calculating the Generated Third Harmonic Electric Field}
In this section, we solve for the generated third harmonic electric field using a Green function approach. When there is no graphene (i.e. in the \emph{bare waveguide}), we can solve for the magnetic field along the propagation direction, $H_z$ using the Helmholtz wave equation for TE modes at $3\omega$, which in rectangular coordinates is given by \cite{Pozar} 
\begin{equation}
( \dfrac{\partial^2}{\partial y^2} + \dfrac{\partial^2}{\partial z^2} + k^2 ) H_z(y,z;3\omega) =0,
\end{equation}
To solve for the current at the graphene, we first consider a \emph{wire} of graphene  located at $ y = \dfrac{b}{2}$ and $z = z_0$, where the current density is given by $J_x^{tot} (y,z_0;3\omega)= J_0^{tot}(z_0; 3\omega) \delta(y- b/2) $. The solutions for the forward and backward propagating fields relative to $z = z_0$ can be expanded exactly as 
\begin{equation}
H_z^{+}(y,z;3\omega)  = \sum_{n=1}^{\infty} B_n \cos(k_n^0 y)e^{i\beta^0_n(3\omega) (z-z_0)},
\end{equation}
and
\begin{equation}
H_z^{-}(y,z;3\omega)  = \sum_{n=1}^{\infty} B_n \cos(k_n^0 y)e^{{-i\beta^0_n(3\omega)} (z-z_0)},
\end{equation}
respectively, where we required $H_z$ to be continuous at $z=z_0$. In these equations we used the fact that for the \emph{bare waveguide} $\tilde{k}_n = k_n^0 = \dfrac{n \pi}{b}$. To evaluate the current, we need to calculate $H_y$, which from Maxwell's equations is given by $H_y =  \dfrac{{i\beta^0_n(3\omega)}}{{k^0_n}^2} \dfrac{\partial H_z}{\partial y}$. Then, the current at this point is related to the $H-$field discontinuity by 
\begin{align}
J_x^{tot}(y,z_0;3\omega) =& [H_y^{-}(y,z;3\omega) - H_y^{+}(y,z;3\omega)]_{z=z_0}\\
=& \sum_{n =1}^{\infty} \dfrac{i\beta^0_n(3\omega)}{{k}^0_n} B_n \sin(k_n^0 y) + \sum_{n =1}^{\infty} \dfrac{i\beta^0_n(3\omega)}{{k}^0_n} B_n \sin(k_n^0 y)\nonumber \\
=&  \sum_{n =1}^{\infty} \dfrac{2i\beta^0_n(3\omega)}{{k}^0_n} B_n \sin(k_n^0 y), \nonumber
\end{align}
We use the Fourier transform to obtain the coefficients, $B_n$: 
\begin{align}
B_n =&  \dfrac{-i k^0_n}{\beta^0_n(3\omega) b} \int_0^b J_0^{tot}(z_0; 3\omega) \delta(y-\dfrac{b}{2}) \sin(k_n^0 y)dy. 
\end{align}
From Maxwell's equations, the generated third harmonic electric field is given by 
\begin{equation}
E_x^{(3)} (y,z;3\omega)=  \dfrac{i 3 \omega \mu}{{k_n^0}^2} \dfrac{\partial H_z(y,z;3\omega)}{\partial y}.
\end{equation}
Using $ H_z (y,z;3\omega) = \sum_{n=1}^{\infty} B_n \cos(k_n^0 y)e^{i\beta^0_n(3\omega)  \vert z - z_0 \vert}$ in Eq. (14), we obtain
\begin{align}
E_x^{(3)} (y,z; 3\omega)=& \sum_{n=1}^{\infty} -\dfrac{i3 \omega \mu}{{k}^0_n} B_n \sin(k_n^0 y) e^{i {\beta}^0_n (3\omega) \vert z-z_0 \vert}\\
=& \sum_{n=1}^{\infty} - \dfrac{3\omega \mu}{ \beta_n^0(3\omega)b} \sin (k_n^0 y) e^{i \beta_n^0 (3\omega)\vert z-z_0 \vert} \int_0^b J_0^{tot}(z_0; 3\omega) \delta(y-\dfrac{b}{2}) \sin(k_n^0 y) dy \nonumber\\
=&  \sum_{n=1}^{\infty} - \dfrac{3\omega \mu}{ \beta_n^0(3\omega)b} \sin(k_n^0 b/2) \sin (k_n^0 y) e^{i \beta_n^0(3\omega) \vert z-z_0 \vert} J_0^{tot} (z_0; 3\omega), \nonumber
\end{align}
The total nonlinear current density at $3 \omega$ in the graphene at $y = b/2$ and $z = z_0$ is given by  
\begin{equation}
J_0^{tot} (z_0; 3\omega) = \sigma^{(3)} (\omega) [E_x^{(1)} (b/2,z_o;\omega)]^3 + \sigma^{(1)} (3 \omega) E_x^{(3)} ( b/2,z_0;3\omega), 
\end{equation}
where, $E_x^{(1)} (b/2, z_0;\omega)$ is the input field for the $n = 1$ mode defined as
\begin{equation}
E_x^{(1)} (b/2, z_0;\omega) = E_{1} e^{i\tilde{\beta}_{1}(\omega)z_0} \sin({k}^0_{1} b/2).
\end{equation}
The generated third harmonic field for a finite length of the graphene sample at the centre of waveguide extends from $0$ to $L$ is 
\begin{equation}
E_x^{(3)} (b/2,z;3\omega) =  \sum_{n = 1}^{\infty} \dfrac{-3 \omega \mu}{{ \beta}^0_n(3\omega)b } \sin(k_n^0 b/2) \sin(k_n^0 y) \int_{0}^{L} J_0^{tot} (z_0; 3\omega) e^{i\beta_n^0(3\omega) \vert z - z_0 \vert} d z_0.
\end{equation}
\section{Linear and Nonlinear Conductivities}
We determine the linear intraband conductivity of the graphene from the Drude-like Model. The intraband linear conductivity of the graphene has been calculated by Horng et al. \cite{horng} It is related to the Fermi energy and the radiation frequency by  
\begin{equation}
\sigma_i^{(1)}(\omega) = \dfrac{v_F e^2 \tau E_f }{\pi \hbar^2(1+ i\omega \tau)}, 
\end{equation}
where $v_F$ is the Fermi velocity of the electrons in the graphene, taken to be $1.1 \times 10^6 \, m/s$, and $\tau $ is phenomenological scattering time which in this work is $50 \, fs$. \\
Following the work done by Al-Naib et al. \cite{Alnaib2015}, the linear interband conductivity is given by
\begin{equation}
\sigma_e^{(1)} (\omega) = -\dfrac{i e^2 v_F}{2 \pi \hbar} \int_{k_f}^{k_{max}} \dfrac{dk}{2 v_F k-\omega-\dfrac{i}{\tau}} \dfrac{\sinh(\beta \hbar v_F k)}{\cosh(\beta E_f)+ \cosh(\beta \hbar v_F k)},
\end{equation} 
where $\beta$ defined as $\dfrac{1}{k_B T}$, where $k_B$ is Boltzman constant. In Fig. S2, we plot the real and imaginary parts of the linear conductivity at $\omega_0$ and $3 \omega_0$ for different Fermi energies. At low Fermi energies, the interband conductivity is essentially given by intraband conductivity, and both of them contribute to the linear conductivity. In the case that Fermi energy, $E_f$ is greater than $20 \, meV$ from the Dirac points, terahertz radiation with, $\omega_0 /2\pi= 2 \, \:THz$ will not be able to promote significant interband transitions, so the dominant contribution of the linear conductivity at higher Fermi energies is intraband conductivity. The linear conductivity increases with Fermi energy, which means that the loss is high for large Fermi energies.   
\begin{figure}[h]
  \centering
  \includegraphics[width=0.8 \columnwidth]{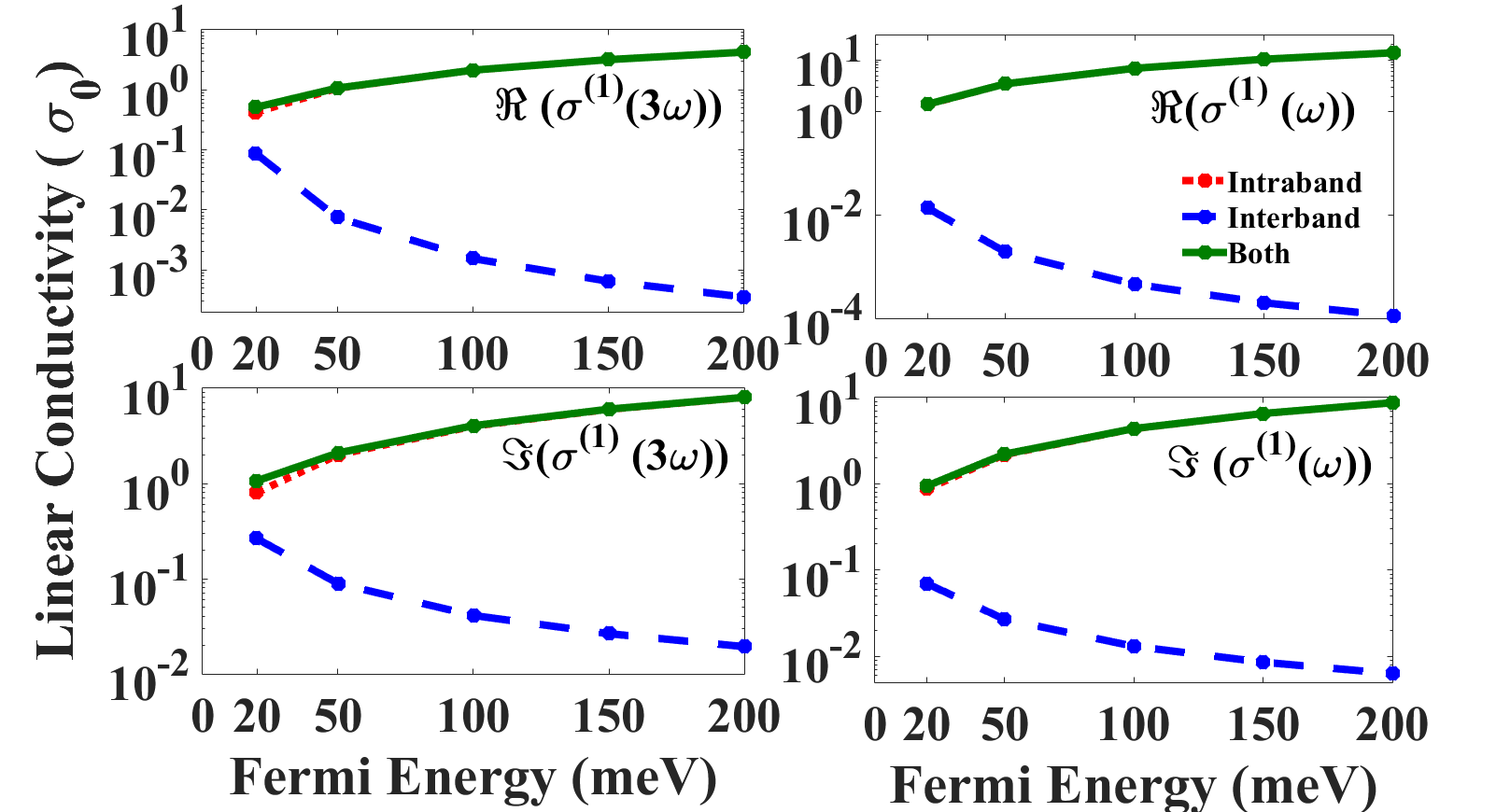}
  \renewcommand{\thefigure}{S\arabic{figure}} 
  \caption{Linear conductivity of the graphene contains interband and intraband conductivities at the frequency of $\omega_0/2\pi = 2 \, \:THz$ and $3\omega_0/2\pi = 6 \, \:THz$ for different Fermi energies in the unit of the universal conductivity of the graphene ($\sigma_0$). }
 \label{fig:Figure3}
\end{figure}\par 
\begin{figure}[h]
  \centering
  \includegraphics[width=0.8 \columnwidth]{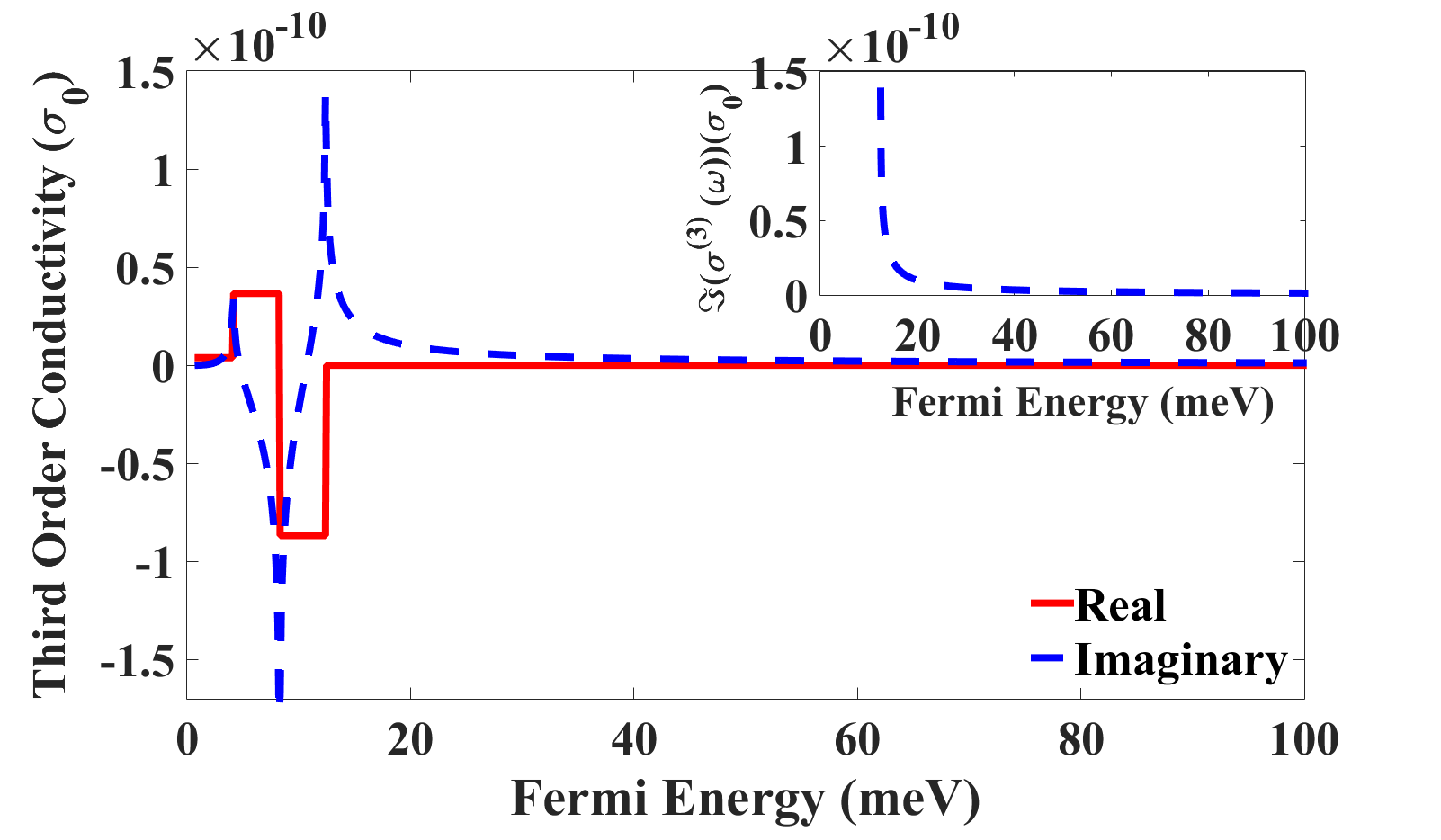}
  \renewcommand{\thefigure}{S\arabic{figure}} 
  \caption{The solid (dashed) curve shows the real (imaginary) part of the third order conductivity. The real part approaches to zero for $E_f > 20 \, meV$, and the imaginary part (shown in the inset) reduces with Fermi energy for $E_f > 20 \, meV$. The fundamental frequency is $\omega_0/2\pi = 2 \, \:THz$ and the conductivity is measured in the unit of the universal conductivity of the graphene, $\sigma_0 = e^2/2 \hbar$.}
   \label{fig:Figure3}
\end{figure}
Sipe et al. in their theoretical work \cite{cheng} explain how third order conductivity depends on the Fermi energy and the frequency of the input field. The third order conductivity of the graphene versus the Fermi energy for a frequency of $\omega_0/2\pi = 2 \, \:THz$ is plotted in Fig. S3. The real part of the third order conductivity tends to zero for $E_f > 20 \, meV$, and the imaginary part decreases with Fermi energy for $E_f > 20 \, meV$, as shown in the inset. Therefore, to have a higher nonlinear conductivity and a lower linear conductivity, so as to potentially higher nonlinear response from graphene, we should work at as low a Fermi energy, as possible.
\section{Numerical Calculations of Third Harmonic Electric Field}
In this section, we show how to calculate numerically the generated third harmonic electric field obtained from Eq. (18). This equation contains two terms, if we replace $J_0^{tot} (z_0;3\omega)$ by Eq. (16). The first integral in the Eq. (18) can be solved analytically. For $ 0 \leqslant z \leqslant L $, for the $n^{th}$ mode, the integration can be evaluated as,
\begin{align}
\int_{0}^{L} \sigma^{(3)} (\omega)  [E_1 e^{ i \tilde{\beta}_{1}(\omega) z_0} \sin(k_n^0 b/2)]^3 & e^{ i{\beta}^0_n (3\omega)\vert z - z_0 \vert} d z_0 \\
= \sigma^{(3)}(\omega) E_1^3 \sin^3(k_n^0 b/2) \bigg \lbrace & \dfrac{e^{i3 \tilde{\beta}_{1}(\omega) z} - e^{i{\beta}^0_n(3\omega) z } }{i 3 \tilde{\beta}_{1}(\omega) - i{\beta}_n^0(3\omega)} \nonumber\\
+& \dfrac{ e^{i3 \tilde{\beta}_{1}(\omega) L} e^{i{\beta}^0_n(3\omega) (L - z)}- e^{i 3 \tilde{\beta}_{1}(\omega) z}}{i 3 \tilde{\beta}_{1}(\omega) + i{\beta}_n^0(3\omega)} \bigg\rbrace. \nonumber
\end{align}
Then, Eq. (18) can be written as
\begin{equation}
E_x^{(3)} (y,z;3\omega) = T(b/2,z;3\omega) + \sigma^{(1)} (3 \omega) \sum_{n = 1}^{\infty} \bigg \lbrace C_n \int_{0}^{L} E_x^{(3)} ( b/2,z_0;3\omega) e^{ i{\beta}_n^0 (3\omega) \vert z - z_0 \vert} d z_0 \bigg \rbrace, 
\end{equation}
where
\begin{align}
T(b/2,z;3\omega) = \sigma^{(3)}(\omega) E_1^3 \sin^3(k_n^0 b/2) \sum_{n=1}^{\infty} C_n & \bigg \lbrace  \dfrac{e^{i3 \tilde{\beta}_{1}(\omega) z} -  e^{i{\beta}^0_n(3\omega) z } }{i 3 \tilde{\beta}_{1}(\omega) - i{\beta}_n^0(3\omega)} \\
&+ \dfrac{ e^{i3 \tilde{\beta}_{1}(\omega) L} e^{i{\beta}^0_n(3\omega) (L - z)}- e^{i 3 \tilde{\beta}_{1}(\omega) z}}{i 3 \tilde{\beta}_{1}(\omega) + i{\beta}_n^0(3\omega)} \bigg \rbrace \nonumber
\end{align}
and 
\begin{equation}
C_n \equiv  \dfrac{- 3 \omega \mu}{{b \beta}_n^0 (3\omega) } \sin(k_n^0 b/2) \sin(k_n^0 y).
\end{equation}
To calculate the third harmonic electric field at the graphene ($y = b/2$) or any other point ($y$), we need to solve Eq. (22) self-consistently. We define the Green function as 
\begin{equation}
G(z, z_0) = \sum_{n=1}^{\infty} C_n e^{- i{\beta}^0_n (3\omega) \vert z - z_0 \vert}
\end{equation}
and, solve Eq. (18) for the generated third harmonic field at the graphene $( y= b/2)$: 	
\begin{align}
E_x^{(3)} (y= b/2,z;3\omega) = T(y=b/2,z;3\omega) + \sigma^{(1)}(3\omega)\int_{0}^{L} G(z, z_0) E_x^{(3)} (y= b/2,z_0;3\omega) d z_0.
\end{align}
Because the Green function diverges when $z=z_0$, to remove this divergence, we add and subtract the following expression to the Eq. (26)
\begin{equation}
\sigma^{(1)} (3\omega) \int_{0}^{L} G(z, z_0) E_x^{(3)} ( b/2,z;3\omega) d z_0.
\end{equation}
Then,
\begin{align}
E_x^{(3)} (b/2,z;3\omega) =& T(b/2,z;3\omega) \\
+& \sigma^{(1)} (3\omega)\int_{0}^{L} G(z,z_0) [E_x^{(3)} ( b/2,z_0;3\omega) - E_x^{(3)} (b/2,z;3\omega)]d z_0 \nonumber\\
+& \sigma^{(1)} (3\omega) E_x^{(3)} (b/2,z;3\omega) \int_{0}^{L} G(z, z_0) d z_0 \nonumber.
\end{align}
To solve Eq. (28) numerically, we discretize $z_0$ and $z$ respectively as $z_{0j} = j \Delta z$, $ z_i = i \Delta z$, where $i,j = 1, ..., N $ and $\Delta z = \dfrac{L}{N-1}$ and switch the integral over $z_0$ to summation. We define,
\begin{align}
G_{ij} =& G(z_i, z_{0j}) \\
K(z_i) =& \int_{0}^{L} G(z_i,z_0) d z_0 = K_i\nonumber
\end{align}
Thus, we obtain
\begin{align}
E_i \equiv & E_x^{(3)} ( b/2, z_i; 3\omega) = T_i + \Delta z \sigma^{(1)} (3\omega) \sum_j G_{ij} [E_j - E_i] + \sigma^{(1)} (3\omega) E_i K_i, \\
E_i = & T_i + \sigma^{(1)} (3\omega) \sum_j \tilde{G}_{ij} [E_j - E_i] + \sigma^{(1)} (3\omega) E_i K_i, \nonumber
\end{align} 
where
\begin{equation}
T_i \equiv T( b/2, z_i; 3\omega)
\end{equation}
and
\begin{equation}
\tilde{G}_{ij} =\left \{ \begin{array}{l}
    \Delta z G_{ij} \quad i \neq j \\
    0 \quad \quad \quad \text{ $i$=$j$}
    \end{array}
    \right.
\end{equation}
Then, we can write Eq. (30) as 
\begin{align}
\sigma^{(1)} (3\omega) \sum_j &\lbrace  \delta_{ij} [1+ p_i -K_i] - \tilde{G}_{ij} \rbrace E_j =  T_i,
\end{align}
where 
\begin{equation}
p_i \equiv \sum_l \tilde{G}_{il}.
\end{equation}
Finally, we define 
\begin{equation}
[F]_{ij} = \sigma^{(1)} (3\omega) \lbrace \delta_{ij} [ 1+ p_i - K_i] - \tilde{G}_{ij} \rbrace,
\end{equation}
and we can solve for the self-consistent third harmonic electric field as, 
\begin{equation}
\mathbf{E} = [\tensor{F}]^{-1} \mathbf{T}.
\end{equation}
We find for our structure that the number of grid points needed to obtain convergence in solving for $E$ is about $N= 1000$.
\section{Dispersion relations}
Fig. S4.(a) shows the effective refractive index differences between $n=1$ mode at $\omega_0$ and $n^{th}$ mode at $3\omega_0$. Two vertical lines are shown on the plot - one at the frequency of $\omega_0/2\pi = 2 \, \:THz$ and one at the $ 3\omega_0/2\pi = 6 \, \:THz$. We choose the fundamental mode to propagate in the $TE_1$ mode, and the horizontal black line shows the three times of the effective refractive index of the first mode at $\omega_0/2\pi$. For the phase matching to occur, one of the modes must come close to the intersection of the horizontal line with the vertical line at $3\omega_0/2\pi = 6 \, THz$; the best phase matching happens between $TE_1$ mode at $\omega_0$ and the $TE_3$ mode at $3\omega_0$. The inset in the Fig. S4.(a) shows the effective refractive index difference for different Fermi energies close to $3\omega_0$. It shows that the best phase matching occurs at low Fermi energies. Fig. S4.(b) represents the loss coefficient for $n=3$ at $3\omega_0$ at different plate separations. In comparison to the loss coefficient for $n=1$ at $\omega_0$, the loss coefficient for $n=3$ at $3\omega_0$ is small; for example, it is $\alpha (3\omega_0) = 0.318 \, mm^{-1}$ for plate separation of $b = 70 \, \mu m$ and $E_f = 50 \, meV$, where for the same structure, $\alpha (\omega_0) = 1.04 \, mm^{-1}$. 
\begin{figure}
  \centering
  \includegraphics[width=0.8 \columnwidth]{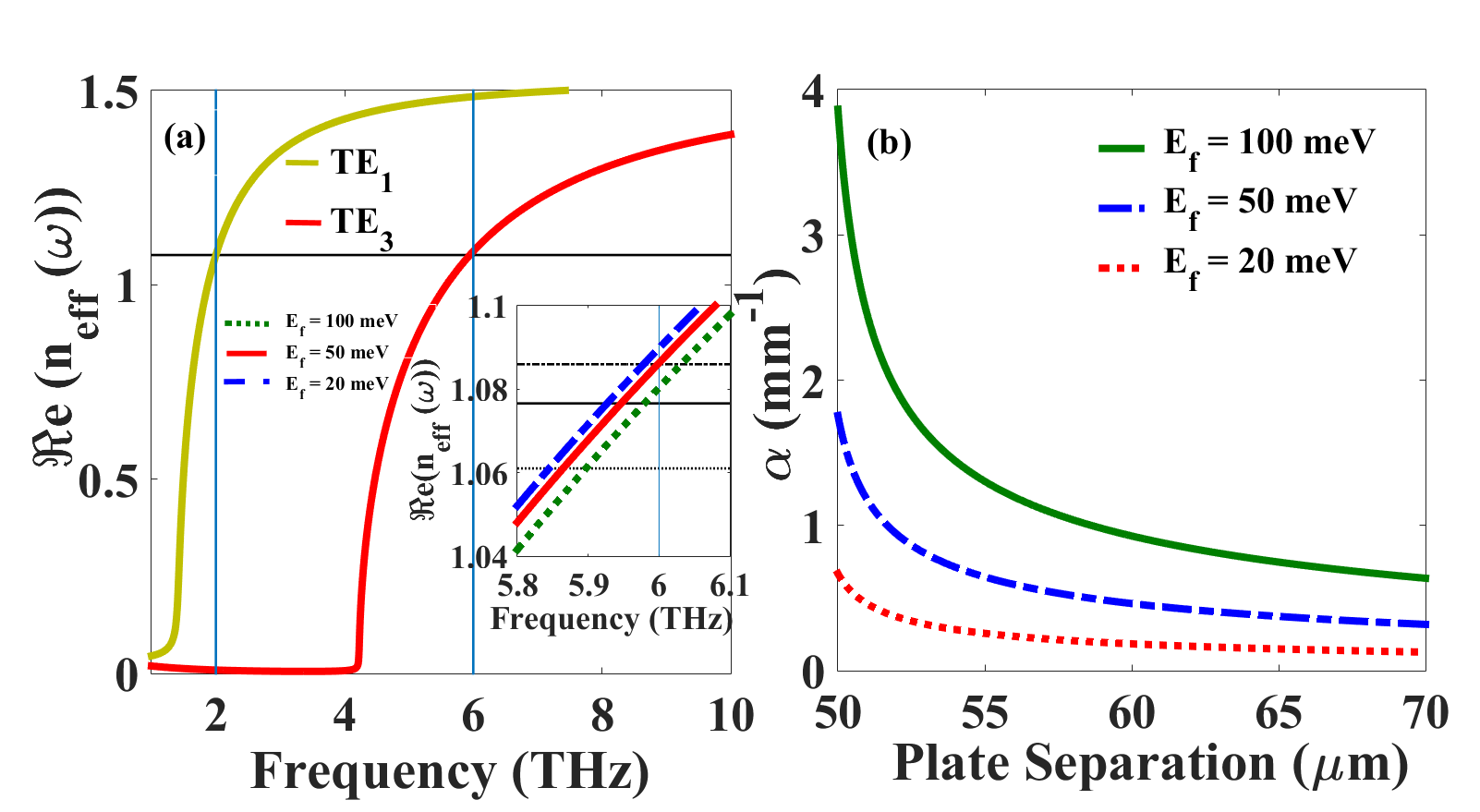} 
  \renewcommand{\thefigure}{S\arabic{figure}}
  \caption{(a): The effective refractive index relation for the waveguide with $b = 70 \, \mu m $ at $E_f = ~50 \, meV$. The first vertical line is at the fundamental frequency $\omega_0/2\pi = 2 \, \:THz$, and the second one is at $3 \omega_0/2\pi = 6 \,\: THz$. The horizontal black line represents three times of the the effective refractive index at $ \omega_0$ for first mode. The effective refractive index relations for different Fermi energies is shown in the inset. (b): Loss coefficient in the $n=3$ mode at $ 3\omega_0$ as a function of plate separation for three different Fermi energies.}
   \label{fig:Figure6}
  \end{figure}
\section{Accurate approximation to evaluate the generated third harmonic field} 
In this section, we derive an approximate method to solve for the generated third harmonic electric field at the graphene with no need to explicitly solve it self-consistently. Using the lossy modes ( replacing $\beta_n^0 (3\omega)$ by $\tilde{\beta}_n (3\omega)$) and including $n=1$ and $n=3$ modes in Eq. (4) in the main text, we obtain 
\begin{align}
G(z,z_0) =& \sum_{n=1,3} \dfrac{- 3 \omega \mu}{2 \tilde\beta_n (3\omega) I}  \sin(\tilde{k}_n^{*} \dfrac{b}{2}) e^{ i \tilde{\beta}_n (3\omega) \vert z - z_0 \vert} dz_0, \nonumber
\end{align}
where $I$ is the normalization factor given by ${\int_0^b \sin(\tilde{k}_ny) \sin(\tilde{k}_n^* y) dy} \simeq b/2 $. 
The total nonlinear current density at $3\omega$ in the graphene in the case of lossy modes is 
\begin{equation}
J_0^{tot} (z;3\omega) = \sigma^{(3)} (\omega) [E_{1}^3 e^{3i\tilde{\beta}_{1} (\omega)z} \sin^3(\tilde{k}_1 \dfrac{b}{2})], 
\end{equation}
since we omit the self-current, which is accounted for approximated in the complex $\tilde{\beta}_n(3\omega)$. Therefore, the generated third harmonic field at the graphene, given by Eq. (3) in the main text, can be written as 
\begin{equation}
E_x^{(3)}(y,z; 3\omega) =  \sum_{n=1, 3}\dfrac{-3 \omega \mu}{2 \tilde{\beta}_n (3\omega) I } \sin(\tilde{k}_n^* \dfrac{b}{2}) \sin(\tilde{k}_n y) \int_{0}^{L}{J}_0^{tot}(z_0;3\omega)  e^{i \tilde{\beta}_n (3\omega) \vert z - z_0 \vert} d z_0, 
\end{equation}
Solving the integral in Eq. (38) analytically, we obtain the generated third harmonic electric field as  
\begin{align}
E_x^{(3)}(y,z;3\omega) = & \sum_{n=1, 3} \dfrac{3i\omega\mu}{\tilde{\beta}_n (3\omega)b} \sin (\tilde{k}_n^{\ast}\dfrac{b}{2})\sin(\tilde{k}_n y) \\
& \times \sigma^{(3)} (\omega) E_1^3 \sin^3 (\tilde{k}_{1} \dfrac{b}{2}) \nonumber \\
& \times \bigg \lbrace \dfrac{e^{i3\tilde{\beta}_{1} (\omega) z} -e^{i\tilde{\beta}_n (3\omega) z}}{3\tilde{\beta}_{1} (\omega) - \tilde{\beta}_n (3\omega)} \nonumber  + \dfrac{  e^{i\tilde{\beta}_{n}(3\omega)(L- z)} e^{i3\tilde{\beta}_{1} (\omega) L} - e^{i3\tilde{\beta}_{1} (\omega)z}}{3\tilde{\beta}_{1} (\omega) + \tilde{\beta}_n (3\omega)} \bigg \rbrace, \nonumber
\end{align}
where we use $I \simeq b/2$. The second term in Eq. (39) is due to the backward propagating wave, which is negligible in comparison to the first term. Calculations for the generated third harmonic electric field by using the approximated method are much easier and faster than the other method explained in the main text and in Sec. \textrm{IV} in the supplemental material.